# An efficient algorithm for three-component key index construction

alexander@veretennikov.ru

In this paper, proximity full-text searches in large text arrays are considered. A search query consists of several words. The search result is a list of documents containing these words. In a modern search system, documents that contain search query words that are near each other are more relevant than documents that do not share this trait. To solve this task, for each word in each indexed document, we need to store a record in the index. In this case, the query search time is proportional to the number of occurrences of the queried words in the indexed documents. Consequently, it is common for search systems to evaluate queries that contain frequently occurring words much more slowly than queries that contain less frequently occurring, ordinary words. For each word in the text, we use additional indexes to store information about nearby words at distances from the given word of less than or equal to *MaxDistance*, which is a parameter. This parameter can take a value of 5, 7, or even more. Three-component key indexes can be created for faster query execution. Previously, we presented the results of experiments showing that when queries contain very frequently occurring words, the average time of the query execution with three-component key indexes is 94.7 times less than that required when using ordinary inverted indexes. In the current work, we describe a new three-component key index building algorithm and demonstrate the correctness of the algorithm. We present the results of experiments creating such an index that is dependent on the value of *MaxDistance*.

In this paper, we continue our research [1]. In the development of modern methods of full-text search, documents that contain queried words near each other are considered more important and relevant [1–4]. The importance of taking proximity information into account in the calculation of relevance increases for larger text collections [3]. At the same time, we need to guarantee that the search time is limited by reasonable boundaries. However, for large text collections, the probability of performance problems related to the search time increases.

Inverted indexes are used for the implementation of the full-text search [5–8]. To take into account the distance between words in the text, we need to store in the index information about every occurrence of every word of every indexed text. Words occur in texts with different frequencies. A typical word frequency distribution in texts [9] (Zipf's law) is presented in Fig. 1. The horizontal axis is used to represent words, with high-frequently occurring words on the left side to low-frequently occurring words on the right side. On the vertical axis, we plot the total number of occurrences in the texts of each word.

The search time is proportional to the total number of occurrences of the queried words in the indexed texts. Therefore, queries that contain high-frequently occurring words can require significantly more time for evaluation (see Fig. 1, on the left side) than queries that consist of ordinary words (see Fig. 1, on the right side). According to [10], the search query, which we consider a "simple inquiry", should be evaluated within two seconds or less. Otherwise, the continuity of the thought of the user could be interrupted, and this interruption could negatively affect the performance of the user. When ordinary inverted files are used, queries that contain high-frequently occurring words often require much more time for evaluation. To solve this performance problem, the author uses additional indexes [1, 11, 12]. Advantages of the proposed approach and a discussion of other methods are presented in [1].

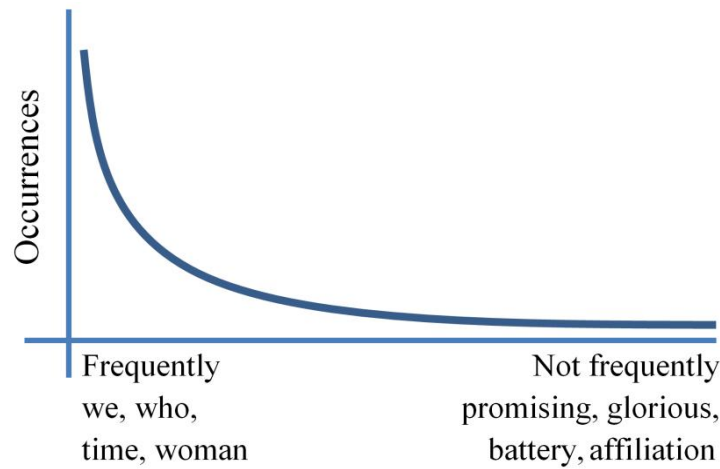

Fig. 1. A typical word frequency distribution

For every word of each document, we can consider a record with two fields (*ID*, *P*), where *ID* is the identifier of the document, for example, its ordinary number within the text collection, and *P* is the position of the word in the document, for example, the ordinary number of the word in the document. Such records are named *postings*. Let us consider two postings *A* and *B*. We define that *A* < *B* when one of the following conditions is met:

1) *A.ID* < *B.ID* or
2) *A.ID* = *B.ID* and *A.P* < *B.P*.

Among the performance improvement methods, the following methods can be considered:

1) Early-termination methods [13, 14] are based on a special sorting of the postings in the index, in order of decreasing the relevance of the posting. At some point in the reading of the posting list, we can decide that the relevance of the current posting is lower than some threshold value, and therefore, the remaining postings can be skipped. It is difficult to integrate the usage of the proximity information into such methods. We always may have some document that contains the queried words near each other, but the same document can also be estimated as a low-relevant document by all other relevance metrics. When we sort the posting list in the index for the specific key, e.g., a word, we cannot take into account all variants of occurrence of all other words near the selected word. See more detailed comments for [2] in [1].
2) Additional indexes methods. In [15], some phrase indexes are presented, but such methods cannot be applied to proximity full-text searches. For this reason, the author proposed the method [1] that allows the solution of the proximity full-text search task.

# § 1. Lemmatization

We use a morphological analyser. The analyser provides a list of basic forms for every word. The basic forms are also named lemmas, and the process of obtaining the list of lemmas for a word is lemmatization. Let *FL*-list be the list of all lemmas. The *FL*-list is ordered in accordance with the occurrence frequencies of lemmas in decreasing order. The three types of lemmas are defined in [11] based on the occurrence frequency of a specific lemma.

Stop lemmas are high-frequently used lemmas, such as: we, who, and time.

Frequently used lemmas often occur in texts and always have a meaning. Examples include book, red.

Ordinary lemmas are all other lemmas, such as promising, glorious, battery, and affiliation.

To divide the set of all lemmas into these three classes, we use the parameters *WsCount* and *FuCount*. In [1], we used *WsCount* = 700 *and FuCount* = 2100.

Let the first *WsCount* elements of the *FL*-list be stop lemmas. Let the following *FuCount* elements of the *FL*-list be frequently used lemmas. Let all remaining lemmas be ordinary lemmas.

In some search engines, stop lemmas can be excluded from the search and the index and, therefore, can be ignored in the search. However, it is stated in [1, 15] that a stop lemma can have a specific meaning in the context of a specific search query in some cases. Therefore, stop lemmas cannot be excluded from consideration, and examples are provided. In our approach, we consider all lemmas.

Let us note that the value of the parameter *WsCount* = 700 is relatively large, and in our definition, stop lemmas can be those that would very rarely be considered stop lemmas in other approaches, such as "time" and "work". In our definition, stop lemmas are those that occur very frequently in texts, and queries that contain such lemmas need a large amount of time for evaluation by means of ordinary indexes.

For *WsCount*, we need to select a value such that, on one hand, all queries that do not contain any stop lemma should be evaluated within the desired time boundaries; on the other hand, the index should be created within acceptable time boundaries. For example, consider the following query: "users need to search". With the numbers in the *FL*-list, we have: user: 307, need: 326, to: 10, search: 70. In the environment that was used in [1], this query can be evaluated within 0.016 sec., when additional indexes are used, and within 70.7 sec., when ordinary inverted indexes are used. The TREC GOV2 text collection was used for this short experiment, and additional indexes were created using *MaxDistance* = 5, *WsCount* = 500, dictionary language – English only. Consider the following query that consists of less frequently occurring lemmas: "sun train manual". With the numbers in the *FL*-list, we have: sun: 635, train: 513, manual: 1296. The latter query can be evaluated by means of additional indexes within 0.004 sec and by means of ordinary inverted indexes within 3 sec. This means that the value of *WsCount* ≈ 500 or 600 can be interpreted as a boundary. Consider a query that consists of lemmas, whose numbers in the *FL*-list are greater than the specified *WsCount*. Such a query can be evaluated within several seconds if ordinary inverted indexes are employed.

*Translation note: the example in the Russian version of the paper is different because it cannot be translated directly.*

In [1], a search methodology is presented that consists of the following points:

1) We divide all lemmas into a specific list of classes based on the occurrence frequency of a lemma: stop lemmas, frequently used lemmas and ordinary lemmas.
2) We use three-component key indexes for the evaluation of queries that consist of only stop lemmas (this is the most complicated case from a performance point of view).
3) We use two-component key indexes for the evaluation of queries that contain a frequently used lemma.
4) We include additional information in the content of posting in one-component key indexes. These indexes are created for frequently used and ordinary lemmas. This information is used for the evaluation of queries that contain a stop lemma and contain a lemma of some other type.

In the current work, we consider the constriction process of three-component key indexes. These indexes solve the subtask of the evaluation of queries that consist only of stop lemmas (the second point of the methodology). Other cases are considered in [11, 12]. We considered the search algorithms in [1]. In the current paper, we consider how the index is built.

## § 2. The algorithm of index construction

### Three-component key index

The extended stop lemma index or three-component key index [1] contains the list of occurrences of the stop lemma *f* for which stop lemmas *s* and *t* both occur in the text at distances that are less than or equal to the *MaxDistance* from *f*. The values of *f*, *s* and *t* are the ordinal numbers of the corresponding lemmas in the *FL*-list. *MaxDistance* is the parameter with example values of 5, 7 or 9.

If the extended (*f*, *s*, *t*) index is constructed, then we can easily produce from this index the indexes (*s*, *t*, *f*), (*t*, *f*, *s*) and indexes with other permutations of *f*, *s* and *t*. Therefore, we build the (*f*, *s*, *t*) index only for the case when $f \leq s \leq t$.

We use multiple execution threads. We create several index files to distribute the construction process over these execution threads. We create an index file for some subset of all three-component key (*f*, *s*, *t*) indexes, and this index file contains the data for the specified three-component key indexes.

The example of the division of the set of all (*f*, *s*, *t*) indexes into several subsets is considered in [16].

The index file is defined by a range of the values of the first component of three-component keys. Moreover, we divide the subset of keys of a specific index file into groups. The group is defined by a range of the values of the second component of three-component keys. This subdivision into groups is performed to optimize the cache usage in the indexing process. Let us consider the following example from [16].

**Example of index file configurations**

**Example 1.** Let *WsCount* be 150. We produce the following index file configuration:

0: [0, 4] → [0, 54] [55, 149];

1: [5, 15] → [5, 32] [33, 60] [61, 104] [105, 149];

2: [16, 52] → [16, 37] [38, 47] [48, 56] [57, 66] [67, 77] [78, 90] [91, 107] [108, 143] [144, 149];

3: [53, 149] → [53, 80] [81, 94] [95, 107] [108, 121] [122, 149].

We have four index files. In the first index file, the range for the first component of keys is [0, 4], in the second [5, 15], in the third [16, 52], and in the fourth [53, 149]. For every index file, we enumerate its groups. Each group is defined by the range of values of the second component of keys. For example, in the index file 0 that is defined by the range [0, 4], the first group is [0, 54] and contains the following keys:

(0, x, y): $0 \leq x \leq 54$, $x \leq y$;

(1, x, y): $1 \leq x \leq 54$, $x \leq y$;

. . .

(4, x, y): $4 \leq x \leq 54$, $x \leq y$.

The data for the keys (5, x, y) are placed into index file 1.

## Construction of the indexes

We use the approach of easily updatable indexes [17]. This approach allows the addition of new data in the index file in several iterations. At each iteration, data of some subset of documents are added into the index. Therefore, the indexing process is the loop in which we repeat the following two-stage process:

1) Read the subset of source documents in sequence one by one. The postings that we read are stored in the RAM in array *D*.
2) Write data from *D* into the indexes.

## Stage 1. Reading of the documents

In the first stage, we read the subset of documents. A document is a sequence of words. For each word, we apply the morphological analyser and produce the list of lemmas *Forms* of the word. For each stop lemma *x* in *Forms,* we produce the record (*ID*, *P*, *FL*(*x*)). In this record, *ID* is the identifier of the document, *P* is the position of the lemma *x* in the document, *FL*(*x*) is the *FL*-number of the lemma *x*, i.e., the ordinal number of x in the *FL*-list. In addition, *Forms* are processed for the construction of indexes of other kinds.

Please note that the (*ID*, *P*, *FL*(*x*)) record requires approximately 3 bytes when it is stored in *D*.

The records that we produced are stored in *D* in RAM. The size of array *D* is limited. When the size of *D* exceeds the limit, we complete Stage 1. If all documents are processed, then we also complete Stage 1.

## Stage 2. Writing the data from *D* into the indexes

Source data: D is the array of records (*ID*, *P*, *Lem*). In this array, *ID* is the identifier of the document, *P* is the position of the lemma *Lem* in the document (ordinal number of the word), *Lem* is the *FL*-number of the lemma. Array *D* is ordered in increasing order of (*ID*, *P*).

For every index file, we start a new execution thread. We limit the number of simultaneously running threads. Let us consider the following example. We limit the number of simultaneously running threads by the number 4, but we have 7 index files. At the start of Stage 1, we create 4 new execution threads for the first four index files. The 5[th] thread will be started only when one of the previously started threads is completed. We designate by Stage 2.1 all work that has been performed in one of the execution threads.

In each execution thread, we organize a loop over the groups of the index file. At each iteration of the loop, we write into the index file the data for the keys ($f$, $s$, $t$), such as the following:

1) $f$ is included in the range of values of the first component of keys that is defined for the index file;
2) $s$ is included in the range of values of the second component of keys that is defined for the group that is in process;
3) $f \leq s \leq t$.

The actions that are needed to process one group of keys are designated Stage 2.1.1. We show the process of indexing for Example 1 in Fig. 2, with the assumption that all execution threads are started simultaneously.

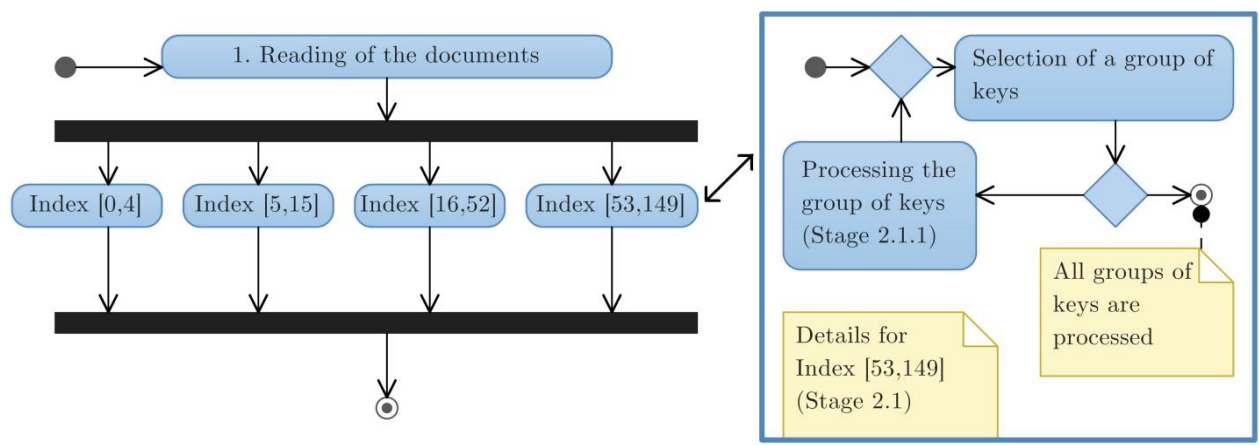

Fig. 2. The two-stage indexing process for Example 1.

## Stage 2.1.1. Processing a group of keys

We read records from array $D$ in sequence, one by one. Let us consider three records from $D$ that correspond with lemmas $f$, $s$ and $t$. We construct a posting for these records if the following conditions are met:

**Condition 1**. Conditions for constructing a posting.

1) $f \leq s \leq t$;
2) ($f$, $s$, $t$) is included in the subset of keys of the current index file, that is, $f$ is included in the range of values of the first component of keys that is defined for the index file;
3) $s$ is included in the range of values of the second component of keys that is defined for the current group;
4) Let us consider the distance between $f$ and $s$ and the distance between $f$ and $t$; the absolute values of these distances must both be less than or equal to *MaxDistance*.

## An algorithm of postings construction and index construction for a group of keys

The main idea of the algorithm is as follows. We iterate over array $D$. For each record ($ID$, $P$) of lemma $f$, we check for the existence nearby of two other lemmas $s$ and $t$ (i.e., both $s$ and $t$ should be at distances that are less than or equal to *MaxDistance* from $f$).

If we found two such lemmas near *f* and (*f*, *s*, *t*) satisfies Condition 1, then we construct posting (*ID*, *P*, *D1*, *D2*), where *D1* is the distance between *f* and *s* and *D2* is the distance between *f* and *t*. Then, we write the posting into the index file for the key (*f*, *s*, *t*).

Let [*IndexS*, *IndexE*] be the range of acceptable values of the first component of keys that is defined for the current index file.

Let [*GroupS*, *GroupE*] be the range of acceptable values of the second component of keys that is defined for the current group.

We use queues. The queue is the data structure that is the singly linked list. The queue also contains two pointers: the start of the queue and the end of the queue. The queue supports the following operations:

1) Add the element to the end of the queue.
2) Remove the element from the start of the queue (the first element).
3) Iterate over elements of the queue from the first element to the last element (from the start to the end). Each element of the queue contains a pointer to the next element; for the last element, this pointer has the value NULL.

## § 3. Simplified algorithm for Stage 2.1.1

Let us create a queue *QueueT*.

In the loop, for every record of *D*, we place this record at the end of *QueueT*.

Let *QueueT.Start* be the start of *QueueT*.

Let *QueueT.End* be the end of *QueueT*.

For every element of the queue, we have a *Processed* flag with an initial value of 0. We set the initial value of the flag when the element is placed into the queue.

Just before the new element (*ID*, *P*, *Lem*) is placed into the queue, we perform the following validation.

Let us consider the following condition: (*P* − *QueueT.Start.P* ) > *MaxDistance* × 2.

While the aforementioned condition is met, we perform the "Extract the first element from the queue" procedure.

The UML diagram of the simplified algorithm for Stage 2.1.1 is presented in Fig. 3.

In steps 2 and 5, queue flushing is performed.

Queue flushing is the process of executing the "Extract the first element from the queue" procedure in the loop until the queue is empty. The queue flushing process is presented in Fig. 4.

Queue flushing is performed in one of the following cases:

1) We move to another document in the process of reading records from *D*.
2) Stage 2.1.1 is completed.

In the UML diagram, we use decision split nodes. Decision split nodes have two forward paths. Each path has a target node. We mark one of these target nodes by a comment. This comment designates the conditions for the corresponding path. If the designated conditions are not met, then another path is selected.

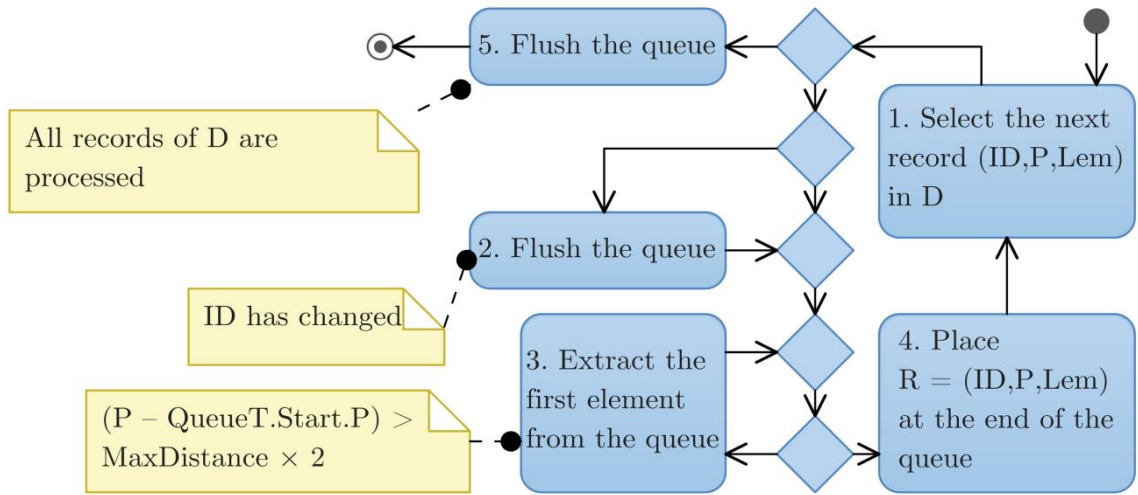

Fig. 3. Simplified algorithm for Stage 2.1.1.

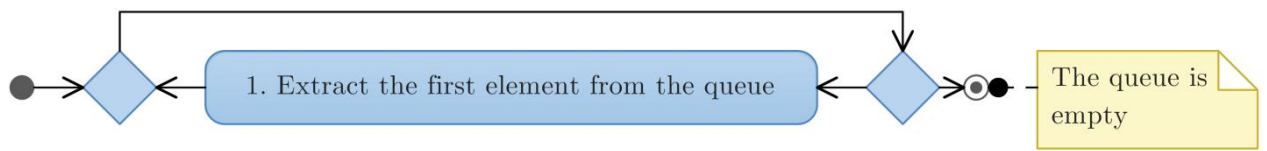

Fig. 4. Queue flushing; see steps 2 and 5 in Fig. 3.

## "Extract the first element from the queue" procedure

In this procedure, we form and write postings to the indexes (steps 1, 2, 3, 4 in Fig. 5). To form a posting, we need to select three elements *F*, *S*, and *T* from the *QueueT* queue. These elements are needed to define three components of (*f*, *s*, *t*) key for the posting, accordingly.

**Condition 2**. Conditions for the selection of the *F* element. This element corresponds to the first component of the key.

*F.P* ≤ *QueueT.Start.P* + *MaxDistance*, *F.Processed* = 0, *IndexS* ≤ *F.Lem* ≤ *IndexE*.

**Condition 3**. Conditions for the selection of the *S* element. This element corresponds to the second component of the key.

|*F.P* – *S.P*| ≤ *MaxDistance*, *F.Lem* ≤ *S.Lem*, *S.P* ≠ *F.P*, *GroupS* ≤ *S.Lem* ≤ *GroupE*.

**Condition 4**. Conditions for the selection of the *T* element. This element corresponds to the third component of the key.

|*F.P* – *T.P*| ≤ *MaxDistance*, *S.Lem* ≤ *T.Lem*, *T.P* ≠ *F.P*, *T.P* ≠ *S.P*.

Let us organize a three-layered loop.

We iterate over all possible elements *F* that satisfy Condition 2.

For the fixed *F*, we iterate in the inner loop over all possible elements *S* that satisfy Condition 3.

For the fixed *F* and *S*, we iterate in the second inner loop over all possible elements *T* that satisfy Condition 4.

For the selected three elements *F*, *S*, and *T*, we form the key (*F.Lem*, *S.Lem*, *T.Lem*).

Then, we form the posting (*F.ID*, *F.P*, *S.P – F.P*, *T.P – F.P*).

Then, we write the posting into the index with the formed key.

The distances between components of the key are included in the posting.

The distances are stored with the sign; therefore, we can determine for both *S.Lem* and *T.Lem* after or before *F.Lem* these lemmas presented in the text.

We also set *F.Processed* = 1.

When the three-layered loop is completed, we remove the first element from the queue.

The procedure is presented in Fig. 5.

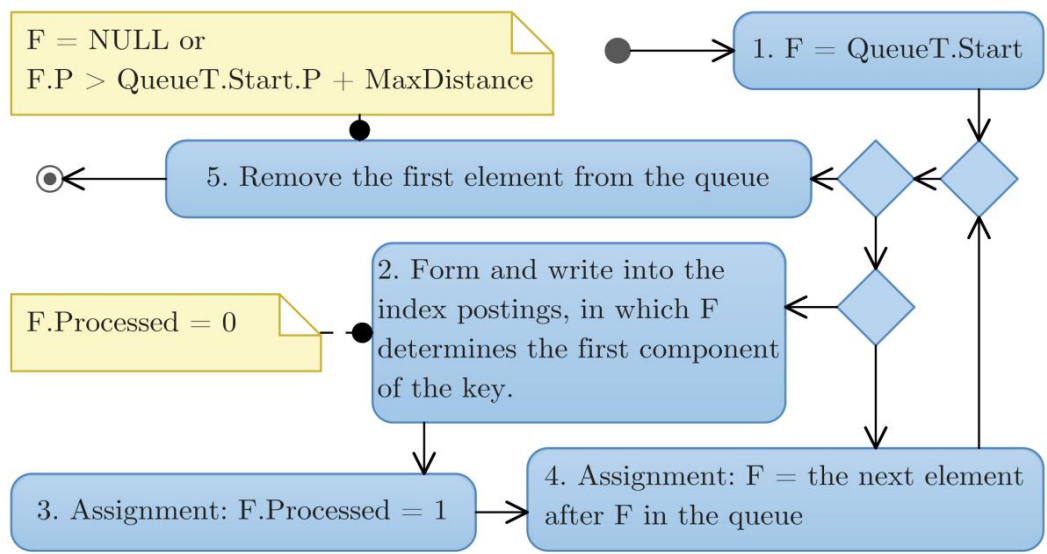

Fig. 5. The "Extract the first element from the queue" procedure for the simplified algorithm. See step 3 in Fig. 3 and step 1 in Fig. 4.

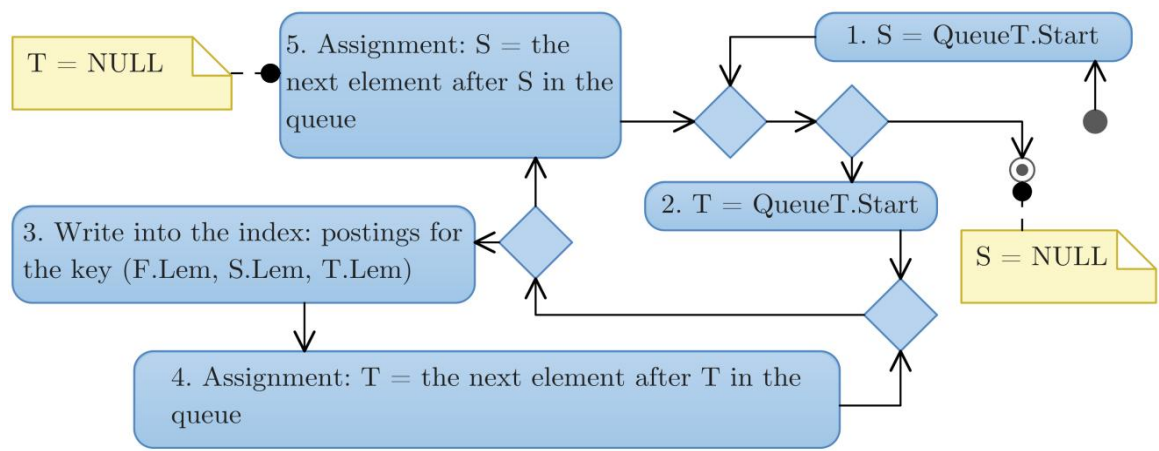

Fig. 6. The "Extract the first element from the queue" procedure for the simplified algorithm; Step 2 in Fig. 5.

### Transition to another document

Only elements with the same value of *ID* should exist in the queue. If in the sequential process of reading *D*, we encounter a record with a different *ID* than all elements of the queue have, we perform the queue flushing. In the queue flushing process, we execute "Extract the first element from the queue" procedure until the queue is not empty. This process is presented in Fig. 4.

After queue flushing, the new element with the new document *ID* is placed in the queue.

## § 4. The optimized algorithm for Stage 2.1.1

We create three queues: *QueueF*, *QueueS*, and *QueueT*. We read records from array *D* in sequence, one by one. The records in *D* are sorted in increased order of (*ID*, *P*).

Let the current record be *R* = (*ID*, *P*, *Lem*).

The purpose of the *QueueF* queue is to store such records *R* that can be used for the first component of the (*f*, *s*, *t*) key.

The purpose of the *QueueS* queue is to store such records *R* that can be used for the second component of the (*f*, *s*, *t*) key.

The purpose of the *QueueT* queue is to store such records *R* that can be used for the third component of the (*f*, *s*, *t*) key.

We skip record *R* if all of the following conditions are met:

1) *R.Lem* is not included in the [*IndexS*, *IndexE*] range.
2) *R.Lem* is not included in the [*GroupS*, *GroupE*] range.
3) *R.Lem* < *GroupS*.

If the aforementioned conditions 1) and 2) are met, then *R.Lem* can only be used as the third component of the key. However, *R.Lem* must be greater or equal to *GroupS*, to be used as the third component of the key.

Otherwise, if the aforementioned conditions are not met for *R*, we perform the following actions.

1) If *R.Lem* is included in the [*IndexS*, *IndexE*] range; then, we place *R* at the end of *QueueF*.
2) If *R.Lem* is included in the [*GroupS*, *GroupE*] range; then, we place *R* at the end of *QueueS*.
3) We place *R* at the end of *QueueT*.

The same record *R* can be placed in several queues.

For every queue, when a new element is placed into the queue, the new element is always greater than or equal to the last element of the queue (in regard to (*ID*,*P*)'s rules of comparison).

In every queue, all elements have the same value of *ID*.

Let *QueueT.Start* be the first element of *QueueT*. Let *QueueT.End* be the last element of *QueueT* (i.e., the last element that was added into the queue).

The following invariant should be preserved:

$$(QueueT.End.P - QueueT.Start.P) \leq MaxDistance \times 2.$$

Just before a new element (*ID*, *P*, *Lem*) is added into the queues, we validate the invariant.

We perform the "Extract the first element from the queue" procedure, while the following condition is met:

$$(P - QueueT.Start.P) > MaxDistance \times 2.$$

## "Extract the first element from the queue" procedure

As a result of this procedure, the first element will be removed from *QueueT*. The first element of the *QueueT* is the minimal element in the *QueueT*. This element will also be removed from *QueueS*, *QueueF*, if possible, i.e., this element exists in some of these queues.

In addition, for each element *F* in *QueueF* with the following condition:

$$F.P \leq QueueT.Start.P + MaxDistance,$$

the following actions will be performed:

1) All postings with the form (*ID*, *F.P*, *X*, *Y*) will be written into the indexes for the keys, where *F.Lem* define the first component of the key (here *X* and *Y* are some distance values that are defined by some other lemmas that are located near *F.P* in the text).
2) *F* will be removed from the *QueueF* queue.

At the start of the procedure, the *QueueT* queue consists of all possible records from *D* with form (*ID*, *P*, *Lem*) that satisfy the following condition:

$$QueueT.Start.P \leq P \leq QueueT.Start.P + MaxDistance \times 2.$$

Let us consider the element *F* from the *QueueF* queue that satisfies the following conditions:

**Condition 5**. Conditions for the selection of the *F* element. This element corresponds to the first component of the key.

1) $F.P \leq QueueT.Start.P + MaxDistance$.

This element will be used for the first component of the key. Element *F* corresponds to an occurrence of a lemma *f* in a text. Lemma *f* will be used for the first component of the (*f*, *s*, *t*) key.

Then, we need to select an element *S* from *QueueS* that will be used to define the second component of the key. This element corresponds to an occurrence of some lemma *s* in texts. Lemma *s* will be used for the second component of the (*f*, *s*, *t*) key.

Let us consider the element *S* from the *QueueS* queue that satisfies the following conditions:

**Condition 6**. Conditions for the selection of the *S* element. This element corresponds to the second component of the key.

1) $S.P \neq F.P$ (different components of the key should correspond to different words with different positions).
2) $S.P \leq F.P + MaxDistance$.
3) $S.Lem \geq F.Lem$ (the condition $f \leq s \leq t$ should be satisfied for the (*f*, *s*, *t*) key).

Then, we need to select an element *T* from the *QueueT* that will be used to define the third component of the key. This element corresponds to an occurrence of some lemma *t* in texts. Lemma *t* will be used for the third component of the (*f*, *s*, *t*) key.

Let us consider the element *T* from *QueueT* that satisfies the following conditions:

**Condition 7**. Conditions for the selection of the *T* element. This element corresponds to the third component of the key.

1) $T.P \neq F.P$, $T.P \neq S.P$.
2) $T.P \leq F.P + MaxDistance$.
3) $T.Lem \geq S.Lem$, $T.Lem \geq F.Lem$.
4) $T.Lem > S.Lem$ or (($T.Lem = S.Lem$) and ($T.P > S.P$)).

For every element *F* of *QueueF*, for every element *S* of *QueueS*, for every element *T* of *QueueT*, which elements satisfy conditions 5, 6, 7, accordingly, we do the following: we save in the index (*F.Lem*, *S.Lem*, *T.Lem*) the new posting (*ID*, *F.P*, *S.P* − *F.P*, *T.P* − *F.P*).

**Theorem 1 (about the correctness of the algorithm)**. When the "Extract the first element from the queue" procedure is executed, for any element *F* from *QueueF*, which element satisfies Condition 5, *QueueT* contains all elements *T* = (*ID*, *P*, *Lem*) of array *D* that satisfy the following condition: $|T.P - F.P| \leq MaxDistance$.

**Proof**. Let us consider an arbitrary record *T* = (*ID*, *P*, *Lem*) of array *D*; this record also satisfies the following condition: $|T.P - F.P| \leq MaxDistance$.

If $T.P \geq F.P$, then $T.P - F.P \leq MaxDistance$; therefore, $T.P \leq MaxDistance + F.P$.

$F.P \leq QueueT.Start.P + MaxDistance$ (Condition 5).

$T.P \leq MaxDistance + QueueT.Start.P + MaxDistance = QueueT.Start.P + MaxDistance \times 2$.

The *QueueT* queue contains all possible records with the latter condition because the "Extract the first element from the queue" procedure is executed in one of the two following cases:

1) Let us consider the record *N* of *D*, where *N* is the next record to read. The following condition is met: $(N.P - QueueT.Start.P) > MaxDistance \times 2$.
2) All records of *D* are processed.

The analysis of case $T.P \geq F.P$ is completed.

Let us now consider the case when $T.P < F.P$.

If $T.P < F.P$, then $F.P - T.P \leq MaxDistance$; therefore, $T.P \geq F.P - MaxDistance$.

If no elements were removed from *QueueT*, then *T* should be in *QueueT*.

Otherwise, let *Q* be the last element that was removed from *QueueT*.

$F.P - Q.P > MaxDistance$ because all elements $F'$ with the condition $F'.P \leq Q.P + MaxDistance$ were processed and removed from *QueueF* in the previous call of the procedure.

$F.P > Q.P + MaxDistance$; therefore, $T.P > Q.P$.

Therefore, *T* should be in *QueueT* because before only elements $T'$ with the following condition were removed from *QueueT*: $T'.P \leq Q.P$. At the same time, $T.P < F.P$ and *QueueT* contains *F*; therefore, *T* was added into *QueueT* beforehand.

The **proof** is completed.

In the optimized algorithm, we do not need the *Processed* flag, and we remove elements from *QueueF* instead using this flag.

**Note 1.** Please note that *QueueS* contains all elements of *QueueT*; these elements can also be used to define the second component of the key. Such elements are placed in both *QueueS* and *QueueT*. We use *QueueS* to optimize the iteration over such elements.

**Note 2.** Please note that the last point in Conditions 7 is used for the exclusion of duplicates.

Let us consider an example. Let us consider lemmas *f* and *s*, $s \geq f$. Let *QueueT* contain exactly one record *F* for lemma *f*. Let *QueueT* contain exactly two records, A and B, for lemma *s*. Let us not take into account the last point of Conditions 7. In this case, two postings can be generated for the key (*f*, *s*, *s*). These are the following postings:

(*ID*, *F.P*, *A.P* − *F.P*, *B.P* − *F.P*) and (*ID*, *F.P*, *B.P* − *F.P*, *A.P* − *F.P*).

In fact, we only need one of them.

**Note 3.** We need to be sure that in all queues, all elements have the same *ID*. This is the same case as in the case of the simplified algorithm. If the new record that we read has an *ID* that is different from the current *ID* that elements in the queues have, then we perform queue flushing. Queue flushing means that we perform the "Extract the first element from the queue" procedure in the loop until all queues are empty.

## Validation by experiments

With Theorem 1, we have a theoretical basis for the algorithm. In addition, we can check the correctness of the indexes by performing search experiments. We can select a document that was indexed. We can produce a list of queries based on the text of the document. Then, we can evaluate each query. For each query, the document and the specific place in the document from which the query was taken should be in the search results. See an example of such an experiment in [1].

# § 5. Key optimizations and the estimation of the performance

## Reconstruction of the array D

The total length and total size of *D* affect the indexing time. For every group, we need to perform iteration over all elements of *D*. We can optimize the iteration process by reconstructing *D* at some points. Let us divide the indexing process into several phases.

Let us consider an index file. The set of keys of the index file is limited by the range of acceptable values of the first component of keys. For example, in Example 1, we have 4 index files. In the first index file, the acceptable range of the first component of keys is [0, 4], in the second index file [5, 15], in the third index file [16, 52] and in the fourth index file [53, 149]. The ranges of the index files are ordered in increasing order.

Let us imagine that the first index file is written. Then, we can exclude from *D* all records (*ID*, *P*, *Lem*) in which *Lem* ≤ 4. For example, for the second index file [5, 15], each component of every key is greater than or equal to 5 (for any (*f*, *s*, *t*) key, we have $f \leq s \leq t$). For the third index file [16, 52], each component of every key is greater than or equal to 16.

Let us imagine that the second index file is also written. Then, we can exclude from *D* all records (*ID*, *P*, *Lem*) in which *Lem* ≤ 15. This exclusion of some records we call reconstruction of *D*.

We do not want to reconstruct *D* after completing the writing of every index file because in this case, we cannot perform indexing of several index files simultaneously in parallel.

In Example 1, we use *WsCount* = 150, which is relatively small. For *WsCount* = 700, we create 79 index files. In the experiments presented, we divide this set of index files that consists of 79 index files into three groups (15, 23, 41). Then, we organize the indexing process into three phases. After each phase, we reconstruct *D* by removing records that are no longer needed. That means, in the first phase, we are writing the first 15 index files in parallel. We wait until all these index files are written. Then, we reconstruct *D*. Following this, we start the next phase in which we are writing the following 23 index files and so on.

### Equalization of the index file processing time

It is important that all indexing threads perform their work at a similar time. However, the indexing time is dependent on the number of records to write. Keys that contain lemmas with a lower value of the *FL*-number have a larger value of records in comparison with the keys that consist of lemmas with a larger value of the *FL*-number. Therefore, for the index files for which the range of acceptable values of the first component of keys are defined by small numbers, the length of the ranges should be less than that for the following index files. For example, consider Example 1. The range for the first index file is [0, 4] and that for the second index file is [5, 15].

However, if *WsCount* is relatively large, then that is not enough. We can use the second component of keys to define the subset of keys of the index file. For example, let us consider Example 1 and the second index file. The index file is defined by the range [5, 15]. For the index file, we define four groups of keys [5, 32], [33, 60], [61, 104], [105, 149], and each group is defined by the acceptable values of the second component of keys.

Instead of creating one index file, we can create two index files. For the first new index file, we use [5, 15] as the range for the first component of keys and use two ranges, [5, 32], [33, 60], for the second component of keys. For the second new index file, we also use [5, 15] as the range for the first component of keys but use two ranges, [61, 104], [105, 149], for the second component of keys.

### Coefficient of utilization

We limit the number of simultaneously running indexing threads. This is because each index thread needs a significant amount of cache that depends on the count of keys of its index file. For each key, we need some amount of memory for the cache. Usually, we have such an amount of index files that we do not have memory to write all of them simultaneously. In this case, we start some amount of index threads. Then, we wait for a thread of them to complete. Then, we can start another index thread and so on. It will be satisfactory, if at the end of the indexing process, we had the maximum number of threads running, and then all of these threads are completed simultaneously.

Let us introduce the *RefCount* variable, this is the number of running threads at a moment in time. When we start a thread, we increment *RefCount*. When a thread is completed, we decrement *RefCount*. Additionally, when *RefCount* needs to be changed, we have *Delta*, which is the time that has passed from the previous change of the value of *RefCount*. Therefore, at each moment of changing *RefCount,* before the change of *RefCount*, we have a record *(RefCount, Delta)* that we push in a list.

Therefore, we have *n* records, $(RefCount_i, Delta_i)$, $1 \leq i \leq n$, each of which corresponds to a time interval of length $Delta_i$ seconds. Here, *n* is the count of the changes of *RefCount,* that is, the count of events of starting or completing a thread. Let us consider the time interval $Delta_i$. During this time interval, we have $RefCount_i$ running threads. Let $MaxRefCount$ be the maximum number of simultaneously running threads during the entire indexing process.

$$MaxRefCount = \max_{1}^{n}(RefCount_i).$$

The entire indexing time is divided into the list of intervals $(RefCount_i, Delta_i)$. This list consists of $n$ intervals.

The entire indexing time is

$$TotalDelta = \sum_{1}^{n} Delta_i.$$

We calculate the utilization coefficient $U$:

$$U = \left(\sum_{1}^{n}(RefCount_i \times Delta_i)\right) / \left(\sum_{1}^{n}(MaxRefCount \times Delta_i)\right).$$

Let us imagine that $U = 1$. In this case, at every moment of time, we had the maximum number of threads running. When the indexing was completed, all threads that were running at this moment completed their work simultaneously.

This can be done, for example, if each indexing thread completed their work at the same time. Moreover, the total number of index threads is divisible by $MaxRefCount$.

Additionally, $U$ can be 1 if we start all of the indexing threads simultaneously and all of them complete their work simultaneously.

If $U$ is near 1, then the computer resources are used in an effective way.

In our experiments, we have $U \geq 0.8$.

The maximum load coefficient $M$ is defined by which part of time we have the maximum number of running threads.

$$M = \left(\sum_{1}^{n}(eq(RefCount_i, MaxRefCount) \times Delta_i)\right) / (TotalDelta), \text{ where}$$

$$eq(a,b) = \begin{cases} 1, & \text{if } a = b, \\ 0, & \text{if } a \neq b. \end{cases}$$

In our experiments, we have $0.55 \leq M \leq 0.8$.

## § 6. Experiments

The following computational resources were used for the experiments.

Intel Xeon X5650 2,67 GHz (2 processors, 6 cores each), 48 GB. RAM.

Windows Server 2008 R2 Enterprise, x64 bit.

HGST HUS726060AL, 6 Tb, SATA.

Text collection from [1] was used.

The text collection contains 195 thousand files with a total size of 71.5 GB. All files are ordinary texts and single byte encoded. The text collection consists of fiction books and magazine articles. The primary language is Russian and queries are also in the Russian language.

We used *WsCount* = 700 and *FuCount* = 2100.

We created three indexes: Idx1 with *MaxDistance* = 5, Idx2 with *MaxDistance* = 7 and Idx3 with *MaxDistance* = 9.

The total sizes of the indexes are as follows:

Idx1: 746 GB., Idx2: 1.23 TB., Idx3: 1.88 TB.

The total building times of the indexes are as follows:

Idx1: 14 h 43 min, Idx2: 19 h 31 min, Idx3: 26 h 49 min.

The total sizes of the three-component key indexes are as follows:

Idx1: 425 GB., Idx2: 883 GB., Idx3: 1.45 TB.

The total times of the three-component key index construction are as follows (see Fig. 7).

Idx1: 8 h 06 min, Idx2: 12 h 39 min, Idx3: 18 h 47 min.

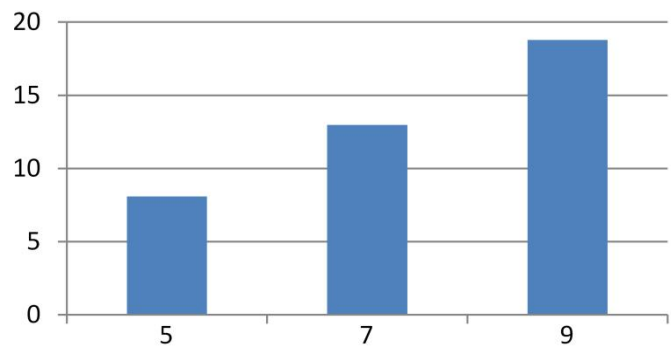

Fig. 7. The total times of the three-component key index construction (hours)

for *MaxDistance* = 5, 7, 9.

## § 7. Additional questions

In the aforementioned experiments, we created several index files, and we used the NTFS file system. The following question arises. Can the particularities of the file system affect the performance? To investigate this question, we created another index with *MaxDistance* = 5. In this experiment, we saved data directly into logical volume.

We created a primitive file system for this goal. In this file system, every file is stored in a list of large blocks. The size of each block is 64 MB. When a file needs to be extended, we allocate a new large block on the volume for this file. The volume is divided into two areas: a data area and free space area. When a new large block needs to be allocated, we allocate it at the start of the free space area. Under these conditions, we created the index at 13 h 53 min. This time is relatively lower than 14 h 43 min, which is required to create a similar index on NTFS volume. The average search times did not change.

Please note that the total sizes of the proposed indexes are significantly greater than the sizes of ordinary inverted indexes. However, the goal of the method is to increase the search speed by several orders of magnitude. In many cases, with the use of the proposed method, the total cost of a search system can be significantly decreased. Frequently, the volume of disk space can more easily be increased than buying several new servers. Additionally, the following question arises. Can the index compression be effectively applied to the proposed indexes?

We performed a preliminary experiment. We used Zip compression. We tried to compress the entire index and tried to compress specified posting lists. The size of the compressed data is approximately 70% of the uncompressed data.

Let us now consider the relevance. Let us consider a search query Q that consists of n words. The search result is the list of records. Each record contains the identifier of a document ID. Each record also contains a list of positions of queried words in the document, that is $X = (X_1, ..., X_n)$.

In [2], the following relevance function is used: $S = \alpha \cdot SR + \beta \cdot IR + \gamma \cdot TP$. Here, SR is the static rank of the document ID, and this value is not dependent on the search query; for example, PageRank can be used. IR can be used for taking into account statistical and other information of queried words, for example, BM25. TP can be used for taking into account proximity information, that is, how near to each other the queried words occur in the document. The values of SR, IR and TP are normalized (i.e., each of them is a number in the range [0, 1]), and $\alpha, \beta, \gamma$ – are parameters.

Often, the value of TP is determined by a number that is inversely proportional to the square of the distance between the queried words in the document. In [2], every query consists of two words and $TP(A,B) = \frac{1}{|A-B|^2}$, where A and B are positions of the queried words in the document (e.g., ordinal numbers of the words). In [18], some values, which are inversely proportional to the square of the distance between queried words in the document, are also used for the calculation of TP. The values are combined with the BM25 value to produce the final rank.

For the case when the query consists of more than two words, we propose the following function.

$$TP(A,B) = \frac{1}{(|A(X)-B(X)|-(n-2))^2}, \text{ where } A(X) = \min_{1 \le i \le n} X_i, B(X) = \max_{1 \le i \le n} X_i.$$

We performed search experiments with queries that consist of frequently occurring words. In these experiments, we used the ordinary inverted index. If we consider a query that consists of frequently occurring words, then we usually have many documents with similar values of BM25 in the result list. Let us introduce the following parameter: *RankBorder* = 0.9. If a query consists of frequently occurring words, then in the search results, we have many documents with normalized BM25 values ≥ *RankBorder*. In extreme cases, when queries consist of high-frequently occurring words, we can have several thousand such documents, each of which has a normalized BM25 value that is greater than or equal to 0.9. Please note that the average document size in the indexed text collection is approximately 54 thousand words. We suppose that if a query consists of frequently occurring words that occur in a large number of documents, then information retrieval ranking functions, such as BM25, do not allow the selection of a short list of relevant documents. Therefore, we require additional relevance criteria, *TP*, for such queries.

When our additional indexes are used, we can find only documents in which the queried words occur near each other, and the distance between the queried words in the text must be no more than *MaxDistance*. Therefore, the value of *MaxDistance* should be large enough, to find all occurrences of queried words in the documents with a large value of *TP*. This can be achieved if the value of *TP* is inversely proportional to the square of the distance between queried words.

For example, let *MaxDistance* be 9. For any query with length ≤ 7, and for any search result $X$ with condition $|A(X) - B(X)| > 9$, we have $TP \leq 1/25 = 0.04$. For example, let us consider a query with a length of 7 words. If a document contains the queried words in the form of a phrase, then $|A(X) - B(X)| = 6$ and $TP = 1$. The value of *TP* will be lower if any "unnecessary" words occur between queried words in the document. For example, if $|A(X) - B(X)| = 10$, then $TP = 1 / (10 - 5)^2 = 0.04$. Therefore, by means of additional indexes, for any query with length ≤ 7, we find in the documents all occurrences of the queried words with a large value of $TP > 0.04$. Longer queries should be divided into parts [1].

Let us imagine that we have some occurrences of queried words in the texts, the value of *SR* is large, and the value of *TP* is small. Such occurrences can be omitted in the search by means of additional indexes. Such occurrences, with small values of *TP*, can be considered to be low relevant and can be skipped. In addition, after the search by means of additional indexes, we can execute an additional search in the ordinary index without considering the distance between words [1]. The search, in which distance between words is not considered, required only a document-level index, the index in which for every word of every document we store only one record. The search in such an index can be performed significantly faster than can be accomplished in the word-level index.

## § 8. Conclusion

We developed an algorithm of three-component key index construction. The results of construction experiments are presented. The following goals are accomplished.

We showed that three component indexes can be created for relatively large values of MaxDistance (i.e., 5, 7, 9).

We showed that with an increase in the value of *MaxDistance*, the total sizes and construction times of the indexes significantly increase.

We showed that the total sizes of indexes that are constructed are significantly larger than the sizes of ordinary inverted indexes, but this factor is not critical because modern data storage devices, such as hard drives, have large capacities.

We determined that the construction algorithm requires many processor resources. The calculation power of the processor is more important than the power of the input/output subsystem for our algorithm.

We proved the correctness of the construction algorithm.

We introduced the utilization coefficient as a criterion for the effectiveness of the usage of computational resources. The value of the utilization coefficient that was measured in the experiments shows that computational resources were used in an effective way.

In the future, it will be interesting to consider ways to optimize the algorithm to allow index construction more quickly. It will also be interesting to determine what values of WsCount and FuCount are optimal, meaning with which values of the parameters could we construct the indexes faster while maintaining the search time within the desired boundaries.

Funding. The original work was supported by Act 211 Government of the Russian Federation, contract № 02.A03.21.0006.

# REFERENCES

1) Veretennikov A.B. Proximity full-text search with response time guarantee by means of three component keys, *Bulletin of the South Ural State University. Series: Computational Mathematics and Software Engineering*, 2018, vol. 7, no. 1, pp. 60–77 (in Russian).
   DOI: 10.14529/cmse180105
2) Yan H., Shi S., Zhang F., Suel T., Wen J.-R. Efficient Term Proximity Search with Term-Pair Indexes, *CIKM '10 Proceedings of the 19th ACM International Conference on Information and Knowledge Management*, Toronto, 2010, pp. 1229–1238.
   DOI: 10.1145/1871437.1871593.
3) Buttcher S., Clarke C., Lushman B. Term proximity scoring for ad-hoc retrieval on very large text collections, *SIGIR '06 Proceedings of the 29th annual international ACM SIGIR conference on Research and development in information retrieval*, 2006, pp. 621–622.
   DOI: 10.1145/1148170.1148285
4) Rasolofo Y., Savoy J. Term Proximity Scoring for Keyword-Based Retrieval Systems, *European Conference on Information Retrieval (ECIR) 2003: Advances in Information Retrieval*, 2003, pp. 207–218.
   DOI: 10.1007/3-540-36618-0_15
5) Zobel J., Moffat A. Inverted Files for Text Search Engines, *ACM Computing Surveys*. 2006. Vol. 38, No. 2. Article 6.
   DOI: 10.1145/1132956.1132959.
6) Tomasic A., Garcia-Molina H., Shoens K. Incremental Updates of Inverted Lists for Text Document Retrieval, *SIGMOD '94 Proceedings of the 1994 ACM SIGMOD International Conference on Management of Data*, Minneapolis, Minnesota, 1994, pp. 289–300.
   DOI: 10.1145/191839.191896.
7) Brown E.W., Callan J.P., Croft W.B. Fast Incremental Indexing for Full-Text Information Retrieval, *VLDB '94 Proceedings of the 20th International Conference on Very Large Data Bases*, Santiago de Chile, Chile, 1994. pp. 192–202.
8) Luk R.W.P. Scalable, statistical storage allocation for extensible inverted file construction, *Journal of Systems and Software archive*, 2011, vol. 84, no. 7. pp. 1082–1088.
   DOI: 10.1016/j.jss.2011.01.049

9) Zipf G. Relative Frequency as a Determinant of Phonetic Change, *Harvard Studies in Classical Philology*, 1929, vol. 40, pp. 1–95.
DOI: 10.2307/408772.
10) Miller R.B. Response Time in Man-Computer Conversational Transactions, *In Proceedings: AFIPS Fall Joint Computer Conference*, San Francisco, California, 1968, vol. 33, pp. 267–277.
DOI: 10.1145/1476589.1476628.
11) Veretennikov A.B. Using additional indexes for fast full-text searching phrases that contains frequently used words, *Control systems and information technologies*, 2013, vol. 52, no 2, pp. 61–66 (in Russian).
12) Veretennikov A.B. Efficient full-text search by means of additional indexes of frequently used words, *Control systems and information technologies*, 2016, vol. 66, no 4, pp. 52–60 (in Russian).
13) Anh V.N., de Kretser O., Moffat A. Vector-Space Ranking with Effective Early Termination, *SIGIR '01 Proceedings of the 24th Annual International ACM SIGIR Conference on Research and Development in Information Retrieval, New Orleans, Louisiana*, USA, 2001, P. 35–42.
DOI: 10.1145/383952.383957.
14) Garcia S., Williams H.E., Cannane A. Access-Ordered Indexes, *ACSC '04 Proceedings of the 27th Australasian Conference on Computer Science, Dunedin*, New Zealand, 2004, P. 7–14.
15) Williams H.E., Zobel J., Bahle D. Fast Phrase Querying with Combined Indexes, *ACM Transactions on Information Systems (TOIS)*, 2004, vol. 22, no. 4, pp. 573–594.
DOI: 10.1145/1028099.1028102.
16) Veretennikov A.B. Efficient full-text proximity search by means of three component keys, *Control systems and information technologies*, 2017, vol. 69, no. 3, pp. 25–32 (in Russian).
17) Veretennikov A.B. About a structure of easy updatable full-text indexes, *Proceedings of the International Youth School-conference «SoProMat-2017»*, Yekaterinburg, Russia, 2017, pp. 30–41 (in Russian).
URL: http://ceur-ws.org/Vol-1894/.
18) Lu X., Moffat A., Culpepper J.S. Efficient and effective higher order proximity modeling, ICTIR '16 *Proceedings of the 2016 ACM International Conference on the Theory of Information Retrieval*, 2016. pp. 21–30.
DOI: 10.1145/2970398.2970404

Veretennikov Alexander Borisovich, Candidate of Physics and Mathematics, Associate Professor, Department of Calculation Mathematics and Computer Science, Ural Federal University,
pr. Lenina, 51, Yekaterinburg, 620083, Russia. E-mail: alexander@veretennikov.ru

See also:

http://www.veretennikov.ru/

http://www.veretennikov.org/Default.aspx?f=Publish%2fDefault.aspx&language=en